# Constrained Nonlinear Model Predictive Control of an MMA Polymerization Process via Evolutionary Optimization

Masoud Abbaszadeh[1] and Reza Solgi[2]

1. Department of Research and Development, Maplesoft, Waterloo, ON, Canada,
2. Swiss Finance Institute, University of Lugano, Via Buffiq 13 6900, Lugano, Switzerland
Email: masoud@ualberta.net, Reza.Solgi@sfi-phd.ch

ABSTRACT

In this work, a nonlinear model predictive controller is developed for a batch polymerization process. The physical model of the process is parameterized along a desired trajectory resulting in a trajectory linearized piecewise model (a multiple linear model bank) and the parameters are identified for an experimental polymerization reactor. Then, a multiple model adaptive predictive controller is designed for thermal trajectory tracking of the MMA polymerization. The input control signal to the process is constrained by the maximum thermal power provided by the heaters. The constrained optimization in the model predictive controller is solved via genetic algorithms to minimize a DMC cost function in each sampling interval.

## 1. INTRODUCTION

Model predictive control (MPC), is a model based advanced process control (APC) technique that has been proved to be very successful in controlling highly complex dynamic systems. It naturally supports design for MIMO and time-delayed systems as well as state/input/output constraints. MPC is generally based on online optimization but in the case of unconstrained linear plants, closed form solutions can be derived analytically. However, when there are constraints over the control inputs (i.e. actuators) and/or process states, which is often the case, an online (i.e. real-time) constrained optimization problem has to be solved in each sampling interval, even if the plant model is linear and time invariant. This online optimization usually requires a high computational power; however, since chemical processes are typically of slow dynamics, such controllers have been designed and implemented on various chemical plants with great success.

Moreover, due to recent advancements of computational hardware and software tools, the usage of MPC is rapidly expanding to other control domains including electrical machines, renewable energy, aerospace and automotive control systems.

In the past two decades, the effective control of polymerization processes control has been studied by many authors [1-7]. Polymerization kinetic is usually complex due to the nonlinearity of the process. Therefore, the control of the polymerization reactor has been staying a challenging task. Due to its great flexibility, a batch reactor is suitable to produce small amounts of special polymers and copolymers. The batch reactor is always dynamic by its nature. It is essential to have a suitable dynamic model of the process. Rafizadeh [1] presented a review on the proposed models and suggested an on-line estimation of some parameters. His model consists of the oil bath, electrical heaters, cooling water coil, and reactor. Peterson *et al.* [2] presented a non-linear predictive strategy for semi batch polymerization of MMA. Soroush and Kravaris [3] applied a Global Linearizing Control (GLC) method to control the reactor temperature. Performance of the GLC for tracking an optimum temperature trajectory was found to be suitable. DeSouza *et al.* [4] studied an expert neural network as an internal model in control of solution polymerization of vinyl estate. In their study, they compared their neural network control with a classic PID controller. Clarke-Pringle and MacGregor [5] studied the temperature control of a semi-batch industrial reactor. They suggested a coupled non-linear strategy and extended Kalman filter method. Mutha *et al.* [6] suggested a non-linear model based control strategy, which includes a new estimator as well as Kalman filter. They conducted experiments in a small reactor for solution polymerization of MMA. Rho *et al.* [7] assumed a first order model plus dead time to pursue the control studies and estimated the parameters of this model by on line ARMAX model. Nonlinear predictive control of the batch reactor considered [8] and [9] via PCA and Wiener modeling approaches, respectively. When MPC is formulated as a state feedback controller, the full state information is required which must be provided using state estimators in nonlinear H2 (e.g. EKF) or nonlinear H∞ paradigms [10-12], [27-28]. Rafizadeh [13] designed a sequential linearization adaptive controller for the solution polymerization of methyl methacrylate in a Batch Reactor.

This paper presents a constrained model predictive control of an MMA reactor, based on the genetic algorithm optimization. A previously developed mechanistic model of the process was used. The model is a sequential piecewise linearization along a selected temperature trajectory. The piecewise linear model is used both for the plant output calculation through the prediction horizon and for the closed loop simulation of the controller, using a time-triggered switching mechanism. We are using an output feedback MPC, therefore, no state estimator is required, which is advantageous. The results of tracking the trajectory and eliminating noise and disturbances show a promising performance of the controller.

## 2. Polymerization Mechanism

Methyl methacrylate normally is produced by a free radical, chain addition polymerization. Free radical polymerization consists of three main reactions: initiation, propagation and termination. Free radicals are formed by the decomposition of initiators. Once formed, these radicals propagate by reacting with surrounding monomers to produce long polymer chains; the active site being shifted to the end of the chain when a new monomer is added. During the propagation, millions of monomers are added to $P_1^o$ radicals. During termination, due to reactions among free radicals, the concentration of radicals decreases. Termination is by combination or disproportionate reactions. With chain transfer reactions to monomer, initiator, solvent, or even polymer, the active free radicals are converted to dead polymer. Table 1 gives the basic free radical polymerization mechanism [14].

The free radical polymerization rate decreases due to reduction of monomer and initiator concentration. However, due to viscosity increase beyond a certain conversion there is a sudden increase in the polymerization rate. This effect is called Trommsdorff, gel, or auto-acceleration effect. For bulk polymerization of Methyl Methacrylate beyond the 20% conversion, reaction rate and molecular weight suddenly increase. In high conversion, because of viscosity increase there is a reduction in termination reaction rate.

TABLE I
POLYMERIZATION MECHANISM

| Initiation | $I \xrightarrow{k_d} 2R^o + G\uparrow$ |
| --- | --- |
|  | $R^o + M \xrightarrow{k_i} P_1^o$ |
|  | $2R^o \xrightarrow{k_{ti}} I'$ |
| Propagation | $P_n^o + M \xrightarrow{k_p} P_{n+1}^o$ |
| Termination | $P_n^o + P_m^o \xrightarrow{k_{tc}} D_{n+m}$ |
|  | $P_n^o + P_m^o \xrightarrow{k_{td}} D_n + D_m$ |
| Transfer | $P_n^o + M \xrightarrow{k_f} P_1^o + D_n$ |
|  | $P_n^o + M \xrightarrow{k_s} s^o + D_n$ |

## 3. MATHEMATICAL MODELING OF POLYMERIZATION

The polymer production is accomplished by a reduction in volume of the mixture. The volumetric reduction factor is given by:

$$\varepsilon = \frac{\rho_p - \rho_m}{\rho_p} \quad (1)$$

The instantaneous volume of mixture is given by:

$$V = \frac{M_0}{\rho_m}(1 - \varepsilon x + \beta) \quad (2)$$

The parameter $\beta$ is defined as:

$$\beta = \frac{f_s}{1 - f_s} \quad (3)$$

During the free radical polymerization, the cage, glass, and gel effects occur. For the cage effect, the initiator efficiency factor is used. The CCS (Chiu, Carrat and Soong) model is used in this study to take into consideration the glass and the gel effects. Therefore, propagation rate constant, $k_p$, is changing according to:

$$\frac{1}{k_p} = \frac{1}{k_{p_0}} + \theta_p \frac{\lambda_0}{D} \quad (4)$$

$k_{p_0}$ is changing as Arrhenius function, and $D$ is given by equation:

$$D = \exp\left[\frac{1 - \varphi_p}{A + B(1 - \varphi_p)}\right] \quad (5)$$

Similarly, termination rate constant, $k_t$, is given by

$$\frac{1}{k_t} = \frac{1}{k_{t_0}} + \theta_t \frac{\lambda_0}{D} \quad (6)$$

$k_{t_0}$ is changing as Arrhenius function. $\theta_p$ and $\theta_t$ are adjustable parameters related to propagation and termination rate constants, respectively. All other necessary parameters and constants for this model are given in [1], [14] and [15].

Long Chain Approximation (LCA) and Quasi Steady State Approximation (QSSA) are used in this study. Equations are highly nonlinear and, using Taylor expansion series, these equations were converted to linearized form. The linearized state space form is given by:

$$\frac{d}{dt}\begin{bmatrix} X \\ i \\ T' \\ T'_j \end{bmatrix} = A \begin{bmatrix} X \\ i \\ T' \\ T'_j \end{bmatrix} + \begin{bmatrix} 0 \\ 0 \\ 0 \\ \frac{2\alpha}{m_o C_{po}} \end{bmatrix} P' \quad (7)$$

$$A = \begin{bmatrix} \frac{\partial f_1}{\partial x}\Big|_s & \frac{\partial f_1}{\partial [I]}\Big|_s & \frac{\partial f_1}{\partial T}\Big|_s & 0 \\ \frac{\partial f_2}{\partial x}\Big|_s & \frac{\partial f_2}{\partial [I]}\Big|_s & \frac{\partial f_{21}}{\partial T}\Big|_s & 0 \\ \frac{\partial f_3}{\partial x}\Big|_s & \frac{\partial f_3}{\partial [I]}\Big|_s & -\frac{UA|_r + UA|_\infty}{mC_p} & \frac{UA|_r}{mC_p} \\ 0 & 0 & \frac{UA|_r}{m_o C_{po}} & -\frac{UA|_r + UA|_{\infty o}}{m_o C_{po}} \end{bmatrix}$$

$$T' = \begin{bmatrix} 0 & 0 & 1 & 0 \end{bmatrix} \begin{bmatrix} X \\ i \\ T' \\ T'_j \end{bmatrix}$$

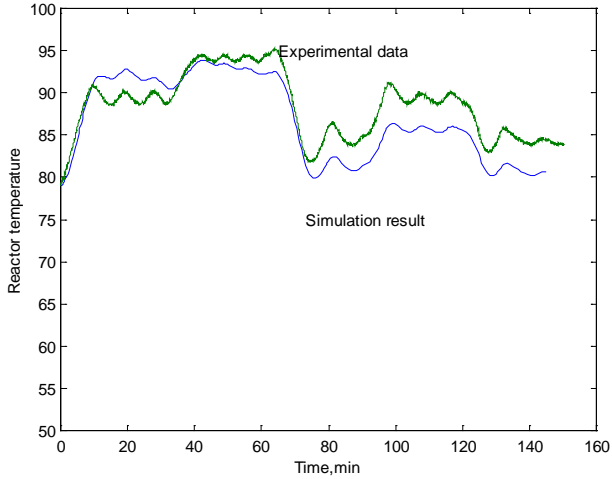

**Fig.1.** Model validation

The molecular properties of the produced polymer are controlled by ensuring the reaction temperature is changing according to a desired reference trajectory. This is a tracking control problem which we are solving using MPC.

Figure 1 shows the result of model validation [14]. As it is seen, the model is a good representative of the process. Equation (7) is converted to the transfer function for reaction temperature to input power:

$$\frac{T'(s)}{P(s)} = \frac{n_3 s^2 + n_4 s + n_5}{d_1 s^4 + d_2 s^3 + d_3 s^2 + d_4 s + d_5} \quad (8)$$

The result of sequential linearization is 131 transfer functions along the temperature profile. See [13] and [14] for a more detailed description of the MMA polymerization dynamic modeling.

## 4. Experimental Setup

A schematic representation of the experimental batch reactor setup is shown in Figure 2 [14]. The reactor is a Buchi type jacketed, cylindrical glass vessel. A multi-paddle agitator mixes the content. Two Resistance Temperature Detectors (RTDs) of PT100 type were used with accuracy of $\pm 0.2\ ^oC$ to measure the reactor temperature and the oil temperature in the oil bath. PT100 sensors provide good linearity in the measurement range and negligible drift. Methyl Methacrylate and Toluene were used as monomer and solvent, respectively. Benzoyl Peroxide (BPO) was used as the initiator. The heater, heats the oil circulating the oil bath, which is pumped into the reactor. Cool water is circulated in a coolant coil inside the oil bath through an electric on/off valve and acts as a safety feature to prevent the oil and consequently the reactor from overheating. The RTD outputs are converted into 0-10 VDC through a bridge and an instrument amplifier and are read by the data acquisition card A/D channels. The controller output is fed into a MOSFET-based power electronics switching circuit as PWM signals. The control command is applied to the MOSFETS through optical isolation, opto-couplers, which provide isolation from a three phase 220V, 50Hz power line. The maximum heater power available is 1000 W, which is a constraint on the control signal.

## 5. Model Predictive Control

Due to its high performance, model predictive control method has received a great deal of attention to control chemical processes, in the last few years. Figure 3 shows block diagram of a model predictive controller.

There are three main approaches to model predictive control, MAC (Model Algorithmic Control), which is based on system's impulse response, DMC (Dynamic Matrix Control), which uses the process step response samples, and GPC (Generalized Predictive Control), which is based on the process transfer function. In practice, it is easier to obtain step response samples rather than impulse response or a full transfer function, and therefore the DMC method is more popular. We use the DMC method in this research. The cost function is defined as:

$$J = \sum_{i=N_1+1}^{N_1+P} \|e(t+i)\|_Q^2 + \sum_{i=1}^{M} \|\Delta u(t+i-1)\|_R^2$$
$$0 \le u(t) \le 1000$$

(9)

where $P$, $M$ and $N_1$ are prediction horizon, control horizon and pure time delay, respectively. Matrices $R_{M \times M}$ and $Q_{P \times P}$ are weighting matrices used in the weighted 2-norms.

$$\begin{aligned} e(t+i) &= y_d(t+i) - y_p(t+i) \\ &= y_d(t+i) - y_m(t+i) - d(t+i) \end{aligned} \quad (10)$$

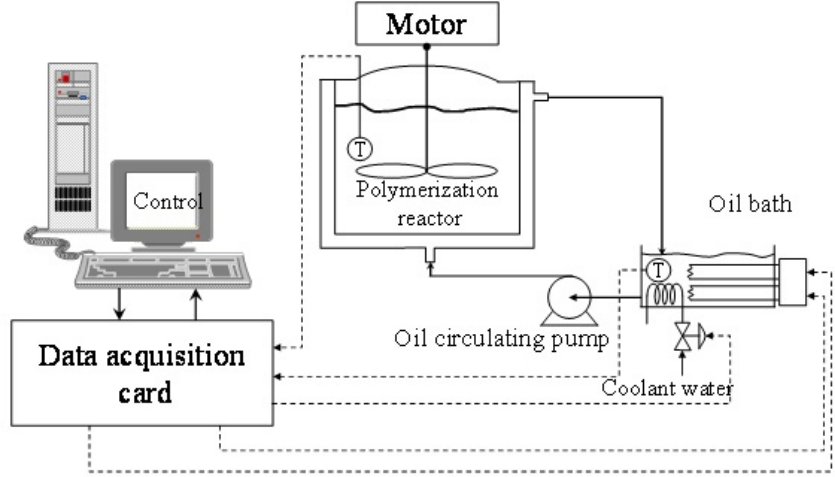

Fig.2. Schematics of the experimental setup

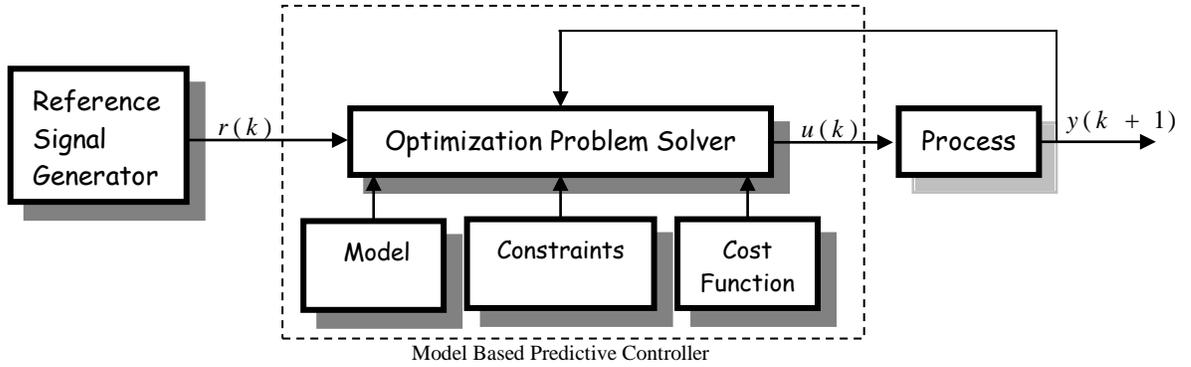

Fig.3. Block diagram of a model predictive controller

where $y_p$ and $y_m$ are the process and model outputs, and $d$ is the difference between the process and model outputs, including noise, disturbance and model mismatch. $y_d(t)$ is the desired output based on the reference input. If $y_{sp}(t)$ is the reference input, the following filtered form is used as the tracking trajectory:

$$y_d(t) = \alpha\, y_d(t-1) + (1-\alpha) y_{sp}(t); \circ \leq \alpha < 1 \quad (11)$$

The parameter α changes the place of the first order smoothing filter pole; the smaller $\alpha$ the faster output will become. It has been shown that system robustness can be decreased by the reduction of $\alpha$ and increment of the manipulated signal [16]. Expanding the summations and substituting the quadratic forms, the cost function in (9) can be rearranged in a matrix form as

$$J = (Y_m + D - Y_D)^T Q (Y_m + D - Y_D) + \Delta U_+^T R \Delta U_+ \quad (12)$$

There is no pure time delay in the model, therefore, $N_1$ is zero, then:

$$Y_m = [y_m(t+1)...y_m(t+P)]^T \quad (13)$$
$$Y_D = [y_d(t+1)...y_d(t+P)]^T \quad (14)$$
$$\Delta U_+ = [\Delta u(t)...\Delta u(t+M-1)]^T \quad (15)$$
$$D = [d(t+1)...d(t+P)]^T \quad (16)$$

In forming (14), if the future values of the set point are known, we get

$$y_d(t+1) = \alpha y_d(t) + (1-\alpha) y_{sp}(t+1)$$
$$y_d(t+2) = \alpha y_d(t+1) + (1-\alpha) y_{sp}(t+2)$$
$$= \alpha^2 y_d(t) + (1-\alpha)[\alpha y_{sp}(t+1) + y_{sp}(t+2)]$$
$$\vdots$$
$$y_d(t+P) = \alpha y_d(t) + (1-\alpha) y_{sp}(t+P)$$
$$= \alpha^P y_d(t) + (1-\alpha) \sum_{i=1}^{P} \alpha^{P-i} y_{sp}(t+i)$$

which is called *programmed MPC*. If the future set points are unknown, $y_d$ is assumed to be constant during the prediction horizon, i.e.:

$$y_d(t+1) = \cdots = y_d(t+P) = y_d(t)$$

which is called *non-programmed MPC*. For an LTI system, without any constraints on output or control signal, the above optimization problem has the following closed form:

$$\Delta U_+ = (G_+^T Q G_+ + R)^{-1} G_+^T Q E \quad (17)$$

where,

$$E \stackrel{\Delta}{=} Y_D - Y_P \quad (18)$$

$$G_+ = \begin{bmatrix} g_\circ & \circ & \cdots & \cdots \\ g_1 & g_\circ & \circ & \\ \vdots & & \ddots & \\ g_{P-1} & & & \ddots \end{bmatrix} \quad (19)$$

and $g_i$ s are the step response samples. The matrix $G_+$ is a Toeplitz matrix consisting the step response samples as shown in (19). There is no closed form solution for the formulated constrained optimization problem. Hence, an online optimization algorithm (a genetic algorithm) is applied to solve the problem, which will be discussed later. The model output is:

$$Y_m = G_+ \Delta U_+ + G_- \Delta U_- + g_N U_N \quad (20)$$

$$G_- = \begin{bmatrix} g_1 & g_2 & \cdots & g_{N-1} \\ g_2 & \cdots & g_{N-1} & \circ \\ \vdots & & & \vdots \\ g_P & \cdots & \cdots & \circ \end{bmatrix} \quad (21)$$

$$\Delta U_- = [\Delta u(t-1) \cdots \Delta u(t-N+1)]^T \quad (22)$$

where $N$ is the number of system step response samples reaching to steady state or equivalent impulse response steps which lead to zero; and $g_N$ is the system DC gain.

$$U_N = [u(t-N) \; u(t-N+1) \; \cdots \; u(t-N+P-1)]^T \quad (23)$$

## 6. THE MODIFIED DMC

If the system has any poles close to the origin, the step response will be very slow and the required $N$ is very large. A system including integrator never reaches to the steady state (this case exists in the set of linearized models of the MMA reactor) and $N$ approaches infinity. Hence, instability occurs. This is one of the limitations of the standard DMC formulation, making it only applicable to open loop stable system [17]. To overcome this, one practical solution is as follows. We have

$$Y_m = G_+ \Delta U_+ + G_- \Delta U_- + g_N U_N \quad (24)$$

$$\overset{\Delta}{Y_{Past}} = G_- \Delta U_- + g_N U_N \Rightarrow Y_m = G_+ \Delta U_+ + Y_{Past} \quad (25)$$

where $Y_{Past}$ is the effect of past inputs on the future system outputs without considering the effect of present and future inputs. Consequently, $Y_{Past}$ can be calculated by setting the future $\Delta u$ s equal to zero and solving the model P steps ahead.

$$\Delta U_+ = \circ \rightarrow Y_m = Y_{Past}$$

As seen in (15) and (19), $G_+$ and $\Delta U_+$ are independent of $N$. The only thing determined by $N$ is the dimension of $G_-$, which is omitted now. Therefore, using this technique, DMC computations become independent of $N$. This modified formulation can be used for marginally stable and unstable plants alike.

As discussed in the previous section, we have used a piecewise linear model of the MMA polymerization process. In our application, since the valid model for future time steps may change through the course of predictions, in the computations of $Y_{Past}$, in order to predict the future model outputs, the corresponding valid models are used. In other words, the model used for output prediction is *scheduled* through the prediction horizon, exploiting the developed trajectory linearized model. This virtual model switching is utilized to calculate the optimal control sequence in each time step. Apparently, another model switching is also applied through the course of time simulation of the closed-loop control system.

Furthermore, since the whole desired temperature profile is known a priori, the programmed MPC is used, utilizing the known future desired temperature trajectory.

## 7. GENETIC ALGORITHM OPTIMIZATION

In general, there is no closed form solution for MPC, except the linear time-invariant unconstrained case. Otherwise, the optimization problem should be solved numerically. If the solution space is convex, sequential quadratic programming techniques (SQP) could be used. Otherwise, either the optimization problem should be a convexified through approximations and relaxation methods or a global optimization algorithm must be used.

Genetic algorithms (GA), are randomized global searching methods developed from the evolution rule in ecological world (the genetic mechanism of survival of the fittest). They have internal implicit parallelism and better optimization ability. By the optimization method of probability, they can automatically obtain and instruct the optimized searching space and adjust the searching direction autonomously [18]. Genetic algorithms, first introduced by Holland [19], are robust global random search methods. These methods are founded based on the Darwinian concept of natural selection and evolution [20], and have been used extensively in optimization and control [21-23].

Coding is essential for the GA optimization. Coding is a mapping from solution space to a finite length strings set. Binary coding is the most generally used method. In this method, each point of the solution space is coded as a binary string, which is a permutation of 0 and 1. Each 0 or 1 is called a *gene* and the string is called a *chromosome*. Every potential solution, $\Delta U_+(t)$, which is a point in the solution space, is coded as a binary word by length of $M \times N_u$. Each element of $\Delta U_+(t)$, is a $N_u$ binary word, called a *subchromosome*. Figure 4 shows the binary coding of the $\Delta U_+$ vector as a chromosome.

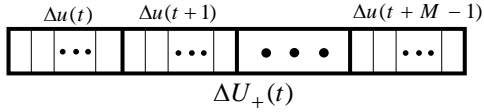

**Fig.4.** Binary coding of $\Delta U_+$ as a chromosome

At the beginning, an initial population that consists of some potential solutions is randomly selected. The final solution is concluded by an iterative method that leads the initial population to an evolutionary one. In each iteration, the next population is generated by applying the crossover and mutation operators to the selected individuals. Their offspring makes the next population. Their selection probability depends on their fitness function that is a measure of *goodness*. The encoding method and fitness function definition are the only links between the physical problem and GA optimization. The following fitness function is applied:

$$F(\Delta U_+) = \frac{1}{1+J} \qquad (26)$$

### a. Encoding
Each element of $\Delta U_+$ is assumed as a 12-binary word. In this way, each chromosome (candidate solution) is a $12 \times M$-binary word.

### b. Selection
In this stage, pairs of the k-th population, $p^k$, are selected to reproduce their offspring. The *tournament selection strategy*, which is a stochastic method, is applied to select each parent. In this method, two individuals are randomly selected and their best (according to their fitness value) is the winner of the tournament, then, the parents are returned to $p^k$.

### c. Crossover
This operator crosses parents to produce the new offspring by gene interchanging. Crossover may occurs in one or more positions in chromosomes. Researchers have suggested several crossover operators such as one point, multipoint and uniform crossover. In this study, a multipoint crossover operator is used to interchange genes between subchromosomes of parents. Figure 5 shows the multipoint crossover genetic operator. The crossover sites are determined randomly.

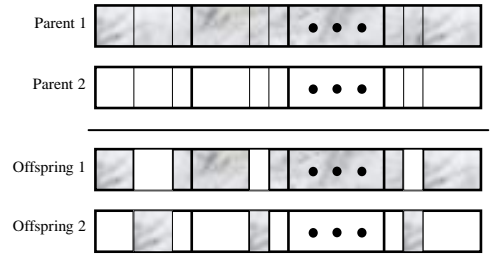

**Fig.5.** Multipoint crossover genetic operator

### d. Mutation
In this phase, a random gene from chromosome is selected and its value will be changed. To do so, a random number between 0 and 1 is generated and compared with the predetermined mutation probability ($P_{mutation} = 0.6$), to see whether the mutation should be done.

### e. Elitist Strategy
Due to stochastic nature of genetic operators, the best individual of $P^{k+1}$ is not essentially better than the $P^k$ one's. A copy of the best individual of current population is directly transferred to the next one to prevent deterioration.

The details of optimization algorithm are depicted in Figure 6. In the controller simulation, the optimization is conducted on a population of 50 chromosomes.

---

1: $k=1$
2: Initialize $P^k$
3: Evaluate fitness value of each chromosome of $P^k$
4: If predetermined termination condition satisfied, **Goto 6**, **Else**
   {
   Insert two copies of the best chromosome of $P^k$ in $P^{k+1}$
   Do for $\frac{N_p}{2} - 1$ times
   {
   Select two chromosomes from $P^k$ as a mating pool
   Apply crossover and mutation to them
   Put their offspring in $P^{k+1}$
   }
   $k := k+1$
   }
5: **Goto 3**
6: Select the best chromosome of $P^k$ as the optimal solution

---

**Fig.6.** Pseudo-code of the applied GA optimization

As mentioned earlier, GA optimization belongs to the family of randomized search algorithms. It is worth mentioning that, in general, there are no theoretical proofs of the speed of convergence of randomized algorithms (convergence within certain time frame, if any). However, since the cost function is convex and the chemical processes are generally slow, allowing to have sampling periods in the order of tens of seconds, the speed of convergence should not be a concern here, provided that the population size and the evolution parameters are selected properly. At the same time, *premature convergence* is avoided by ensuring good genetic variation. The genetic variation can be regained by using a large enough population size and also by mutation [25]. On a separate note, in fast systems with millisecond sampling times, the optimization algorithm may not converge within the sampling interval and the optimization computation might be halted. In such cases, especially in mission-critical and safety-critical applications, hybrid algorithms must be utilized using FSQP (Feasible SQP) solvers, to ensure feasibility (satisfaction of all constraints) of the optimizer in each and every iteration even before convergence.

## 8. SIMULATION RESULTS

The population is the foundation of evolution of a genetic algorithm. The character of the population decides the search capability of the genetic algorithm. The astringency of the genetic algorithm is determined by the astringency of the population [24].

First, the effects of population size and the number of generations on controller performance are studied. As it is shown in the Figure 7, after some thresholds of population size and number of generations, there is no significant improvement in controller performance, in comparison with the great growth deal of computations. The figure exhibits convergence of the GA-based optimization. The sampling period is 30 seconds and other parameters are as follows:

$$P = M = 3, Q = I_{3\times 3}, R = .05 \times I_{3\times 3}, \alpha = .1, N_u = 12$$

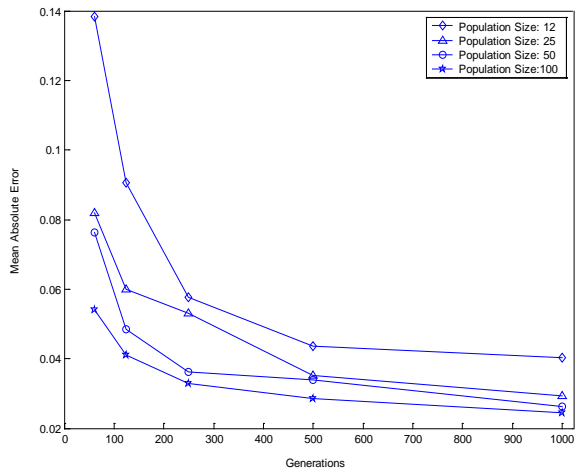

**Fig.7.** Effect of the number of generations and population size on controller performance

Our several simulations show that the GA optimization algorithm converges well within the selected MPC sampling period in all runs.

Figure 8 shows the simulation results of controller performance with a population size of 50 and the number of generations equal to 750. The top plot shows the systems output verses the desired thermal trajectory and exhibits controller's high performance. The middle plot is the control signal, which as seen in the plot, satisfies the constraint. The bottom plot is the tracking error, i.e. the difference between the actual and desired outputs. The DMC controller provides integral action, capable of rejecting step disturbance. The ability of controller to reject noise and disturbance is shown in Figure 9, in which a step output disturbance and a zero mean white Gaussian measurement noise with a variance of $0.1^oC$ (STD ~ $0.3^oC$) are applied. As seen, the controller has good disturbance rejection performance and the average absolute tracking error is less than $0.25^oC$. Compared to the previous results, the adaptive PI control in [13] has a *2° C* average error and the Generalized Takagi-Sugeno-Kang fuzzy controller proposed in [26] has a *1°C* average error; demonstrating the superior performance of the MPC.

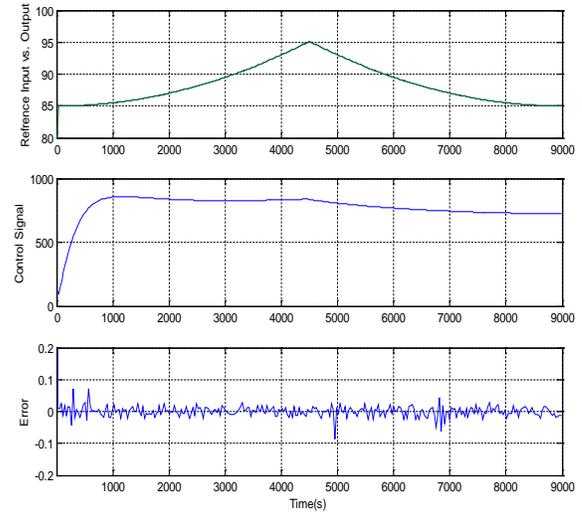

**Fig.8.** The output, control signal and error

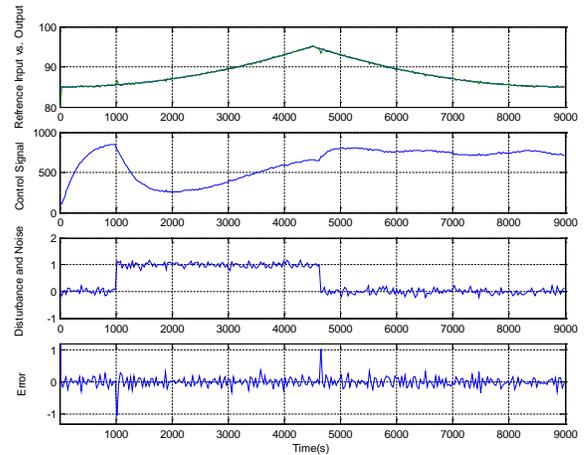

**Fig.9.** The controller performance in the presence of noise and disturbance

## 9. CONCLUSION

A sequential piecewise linearized model based predictive controller based on the DMC algorithm was designed to control the temperature of a batch MMA polymerization reactor. Using the mechanistic model of the polymerization, a transfer function was derived to relate the reactor temperature to the power of the heaters. The coefficients of the transfer function were calculated along the selected temperature trajectory by sequential linearization. A genetic algorithm (GA) is applied for the cost function optimization DMC. The simulation result of controller performance shows that the tracking the profile, noise and disturbances rejection is very good.

## ACKNOWLEDGEMENT
The authors would like to thank Dr. Mehdi Rafizadeh for providing the model of the reactor.

## REFERENCES

[1] M. Rafizadeh, "Non-Isothermal Modeling of Solution Polymerization of Methyl Methacrylate for Control Purposes", *Iranian Polymer J.*, 2001, 10 (4), pp. 251-263.

[2] T. Peterson, E. Hernandez, Y. Arkun and F. J. Schork, "Non-linear Predictive Control of a Semibatch Polymerization Reactor by an Extended DMC", *American Control Conference*, 1989, pp. 1534-1539.

[3] M. Soroush and C. Kravaris, "Non-linear Control of a Batch Polymerization Reactor: An Experimental Study", *AICHE J.*, 1991, 38 (9), pp. 1429-1448.

[4] M. B. De Souza Jr., J. C. Pinto and E. L. Lima, "Control of a Chaotic Polymerization Reactor: A Neural Network Based Model Predictive Approach", *Polym. Eng. Sci.*, 1996, 36 (4), pp. 448-457.

[5] T. Clarke-Pringle and J. F. MacGregor, "Non-linear Adaptive Temperature Control of Multi-Product, Semi-Batch Polymerization Reactors", *Computers Chem. Engng.*, 1997, 21, pp.1395-1409.

[6] R. K. Mutha, W. R. Cluett and A. Penlidis, "On-Line Non-linear Model-Based Estimation and Control of a Polymer Reactor", *AICHE J.*, 1997, 43 (11), pp. 3042-3058.

[7] H. R. Rho, Y. Huh and H. Rhee, "Application of Adaptive Model-Predictive Control to a Batch MMA Polymerization Reactor", *Chem. Eng. Sci.*, 1998, 53 (21), pp. 3728-3739.

[8] J. Flores-Cerrillo, J.F. MacGregor, "Latent Variable MPC for Trajectory Tracking in Batch Processes", *Journal of Process Control*, 15 (6), 2005, pp. 651–663.

[9] G. Shafiee, M.R. Jahed-Motlagh, A.A. Jalali "Nonlinear predictive control of a polymerization reactor based on piecewise linear Wiener model", *Chemical Engineering Journal*, 143 (1–3), 2008, Pages 282–292.

[10] S.M. Ahn, M.J. Park, H.K. Rhee, "Extended Kalman filter-based nonlinear model predictive control for a continuous RIMA polymerization reactor", *Industrial & Engineering Chemistry Research*, 38 (10), 1999, pp. 3942-3949.

[11] M. Abbaszadeh, H.J. Marquez, "Dynamical robust H∞ filtering for nonlinear uncertain systems: An LMI approach", *Journal of the Franklin Institute* 347 (7), 2010, pp. 1227-1241.

[12] M. Abbaszadeh, H.J. Marquez, "Robust static output feedback stabilization of discrete-time nonlinear uncertain systems with H∞ performance", *Proceedings of the 16th Mediterranean Conference on Control and Automation*, 2008, pp. 226-231

[13] M. Rafizadeh, "Sequential Linearization Adaptive Control of Solution Polymerization of Methyl Methacrylate in a Batch Reactor", *Polymer Reaction Eng. J.*, 2002, 3 (10), pp. 121-133.

[14] M. Rafizadeh and M. Abbaszadeh, "Nonlinear Model Predictive Control of Discrete Methyl Methacrylate Polymerization for Temperature Profile Tracking", *Amirkabir Journal of Science and Technology*, 14, (55-C), pp. 811-824, 2003.

[15] M. Abbaszadeh, "Nonlinear Multiple Model Predictive Control of Solution Polymerization of Methyl Methacrylate", *Intelligent Control and Automation*, 2 (3), pp. 226-232, 2011.

[16] E. F. Camacho and C. Bordons, *Model Predictive Control in the Process Industry*, Springer and Verlag, 1995.

[17] P. Lundstorm, J. H. Lee, M. Morari and S. Skogestod, "Limitations of Dynamic Matrix Control", *Computers and Chemical Engineering*, 19 (4), 1995, pp. 402-421.

[18] J. Ho, J. Gu, T. Zhao, G. Sun, "A New Resource Scheduling Strategy Based on Genetic Algorithm in Cloud Computing Environment", *Journal Of Computers,* 7 (1), 2012, pp. 42-52.

[19] J. H. Holland, *Adaptation in Natural and Artificial Systems*, Ann Arbor, MI, University of Michigan Press, 1975.

[20] L. M. Schmitt, "Fundamental Study, Theory of Genetic Algorithms", *Theoretical Computer Science*, 259(2001), pp. 1-61.

[21] K.S Tang, K. F. Man, S. Kwong and Q. He, "Genetic Algorithms and Their Applications", *IEEE Signal Processing Magazine*, November 1996, pp. 22-37.

[22] R. Bandyopadhyay, U. K. Chakaraborty and D. Patranabis, "Autotuning a PID Controller: A Fuzzy-Genetic Approach", *J. of System Architecture*, 47 (2001), pp. 663-673.

[23] L. Bernal-Harro, C. Azzaro-Pantel and S. Domenech, "Design of Multipurpose Batch Chemical Plants Using a Genetic Algorithm", *Computers and Chemical Engineering*, 22 (1998), pp. S777-S780.

[24] Q. Zhio, L. Xi, "Application of Genetic Algorithm to Huasdorff Measure Estimation", *Proceedings of the 11th International Conference of Knowledge-Based Intelligent Information and Engineering Systems,* 2007, pp. 182-189.

[25] Z. Michalewicz, "*Genetic Algorithms + Data Structures = Evolution Programs",* 3rd Edition. Springer-Verlag. 1996.

[26] Solgi R.; Vosough R.; Rafizadeh M., "Adaptive Fuzzy Control Of MMA Batch Polymerization Reactor Based On Fuzzy Trajectory Definition", *Proceeding of the 2004 American Control Conference*, pp. 1097-1102, 2004.

[27] M. Abbaszadeh, H.J. Marquez, "Robust $H\infty$ observer design for sampled-data Lipschitz nonlinear systems with exact and Euler approximate models", *Automatica*, 44 (3), 2008, pp. 799-806.

[28] M. Abbaszadeh, H.J. Marquez, "LMI optimization approach to robust $H\infty$ observer design and static output feedback stabilization for discrete-time nonlinear uncertain systems", *International Journal of Robust and Nonlinear Control*, 19 (3), 2009, pp. 313-340.